\documentclass[preprint,aps,amssymb]{revtex4} 
\usepackage{graphicx}%
\newcommand{\bm}[1]{\mbox{\boldmath$#1$}}

\begin{document}

\title{Marginalization using the metric of the likelihood}

\author{R.\ PREUSS}
\email{preuss@ipp.mpg.de}
\author{V.\ DOSE}
\affiliation{Max-Planck-Institut f\"ur Plasmaphysik, EURATOM Association\\
Boltzmannstr.\ 2, D-85748 Garching b. M\"unchen, Germany}

\date{\today}

\begin{abstract}
Although the likelihood function is normalizeable with respect to the data
there is no guarantee that the same holds with respect to the
model parameters.
This may lead to singularities in the expectation value integral
of these parameters, especially if the prior information is not sufficient
to take care of finite integral values.
However, the problem may be solved by obeying the correct Riemannian metric
imposed by the likelihood.
This will be demonstrated for the example of the electron temperature
evaluation in hydrogen plasmas.
\end{abstract}

\maketitle

\section{Introduction}

Given data $\vec d$, a linear parameter $c$ and some function $\vec f$
meant to explain the data, we have
\begin{equation}
\vec d = c \cdot \vec f ( T ) + \vec\varepsilon
\label{data}
\qquad.
\end{equation}
The vectors shall have dimension $N$ according to the number of
quantities measured.
Due to the measurement process the data is corrupted by noise, where
$\langle \varepsilon \rangle =0$ and $\langle \varepsilon^2 \rangle =
\sigma^2$.
Then by the principle of Maximum Entropy the likelihood function reads
\begin{equation}
p ( D | c , \sigma , \vec f , I ) \propto \exp \left\{ - \frac{1}{2 \sigma^2}
\sum_i [ d_i - c f_i ]^2 \right\}
\qquad,
\end{equation}
which is clearly normalizeable for the data $\vec d$ and bound for every
parameter showing up as a functional dependency in $f$.
The situation may change when we are looking for the expectation value of
some parameter of $f$, let say $f=f(T)$.
Then we need to evaluate the posterior of $T$ with
\begin{equation}
\left\langle T \right\rangle \propto \int \ T \ p ( {\rm d}T | D , I )
\qquad.
\end{equation}
In order to connect the unknown posterior to the known likelihood
we marginalize over all the parameters which enter the problem, that
is in our problem $c$ and $\sigma$:
\begin{equation}
p ( {\rm d}T | D , I )=
\int_c \int_\sigma p ( {\rm d}T , {\rm d}c , {\rm d}\sigma | D , I )
\qquad,
\label{marg}
\end{equation}
and make use of Bayes theorem:
\begin{equation}
p ( T , c , \sigma | D , I ) \propto
p ( D | T , c , \sigma , I ) \  p ( T , c , \sigma | I )
\qquad.
\end{equation}
Commonly, the infinitesimal elements in equation (\ref{marg}) are identified with
\begin{equation}
p ( {\rm d}T , {\rm d}c , {\rm d}\sigma | D , I ) = p ( T , c , \sigma | D , I )\
{\rm d}T \ {\rm d}c \ {\rm d}\sigma
\qquad.
\label{inf}
\end{equation}
In mathematical terms this would mean that the probability functions live in
euclidean space.
They do not.

\section{Riemannian Metric}

Parameterizations correspond to choices of coordinate systems.
The problem to be solved has to be invariant against reparametrizations 
\cite{rod89}, i.e.\ in the space of the probability functions one
has to get the same answer no matter what parameters were chosen to
describe a model.
Therefore one is in need of a length measure $\mu$ which takes care of
defining a distance between different elements of this probability
function space.
This task is done by applying differential geometry to statistical
models, an approach which was baptized 'information geometry'
\index{information geometry} by S.\ Amari \cite{ama85}.
Eq.\ (\ref{inf}) then reads correctly
\begin{equation}
p ( {\rm d}T , {\rm d}c , {\rm d}\sigma | D , I ) = p ( T , c , \sigma | D , I )\
\mu ({\rm d}T , {\rm d}c , {\rm d}\sigma)
\qquad.
\end{equation}
$\mu ({\rm d}\vec \theta)=\mu ( \vec\theta ) {\rm d}\vec \theta$ is the
natural Riemannian metric \index{Riemannian metric}
 on a regular model
(in our case the model is parameterized by $\vec\theta = ( T , c , \sigma
)$).
It results from second variations of the entropy \cite{ama85,rod91} and
is given by
\begin{equation}
\mu ({\rm d}\vec \theta) = \sqrt{ \det \bm{g}(\vec\theta) } {\rm d}\vec \theta
\end{equation}
where \bm{g} is the Fisher information matrix:
\begin{equation}
g_{ij} = - \left\langle \frac{\partial^2 \log p( D | \vec \theta , I)}
{\partial \theta_i \partial \theta_j} \right\rangle
\qquad.
\end{equation}
For the above likelihood the metric reads explicitly
\begin{equation}
\mu (\sigma, c , T) \propto \frac{c}{\sigma^3}
\sqrt{
 \left[ \sum_i f_i^2 \right] 
 \left[ \sum_i \left( \frac{\partial f_i}{\partial T} \right)^2 \right] 
-
 \left[ \sum_i f_i \frac{\partial f_i}{\partial T} \right]^2
}
\qquad.
\label{metric}
\end{equation}
Notice that this approach is based on the assumption that the hypothesis
space of the likelihood defines the metric to be calculated in.
This may not be the case if some prior information was already used
during data acquisition, e.g.\ the experimentalist uses his expert
knowledge in separating 'correct' data from the rest.
The latter instantly rules out certain parts of all possible realizations
of the likelihood function and results in a different hypothesis space.

\section{Simple Example}

First we want to demonstrate the relevance of using the correct metric
with a simple example which already has all the features of the real
world problem further down.
\begin{equation}
f_i ( T ) = T \cdot ( T + x_i )^{-1} x_i
\qquad,
\end{equation}
where the notation in $i$ corresponds to the data points $d_i$.
For simplification let us assume that the variance $\sigma^2$ is known and
we only have to marginalize over $c$ in order to get the posterior.
What happens if we do not use the Riemannian metric?
Then the marginalization \index{marginalization} integral over $c$ reads
\begin{equation}
p ( T | D , I ) \propto \int {\rm d}c \ p ( D | T , c , I ) \ p ( c | I ) 
\qquad.
\end{equation}
In order to facilitate analytic calculation the exponent of the
likelihood is written in a quadratic form over $c$
\begin{equation}
\sum_i [ d_i - c f_i ]^2
=
(\vec f^T \vec f) [ c - c_0 ]^2 +
\left[ \vec d^T \vec d - \frac{( \vec d^T \vec f )^2}
{\vec f^T \vec f} \right]
\qquad,
\end{equation}
where $c_0 = \vec d^T \vec f / \vec f^T \vec f$.
For the prior $ p ( c | I ) $ the only thing we know is that $c$ will be
something in between an upper and a lower limit, where it is
reasonable to assume that the upper (lower) bound is given by an unknown
factor $n$ (1/$n$) of the value $c_0$ where the maximum of the likelihood
occurs.
The principle of maximum entropy gives a flat prior with
\begin{equation}
p ( c | I ) = \left\{
\begin{array}{cl}
\frac{1}{nc_0} & \forall \ 0 \le c \le n c_0 \cr
0                            & {\rm else} \qquad \qquad.
\end{array}
\right.
\label{cprior}
\end{equation}
The integral over the c-dependent parts then reads
\begin{equation}
\frac{1}{nc_0}
\int_{0}^{n c_0}
{\rm d}c \
\exp\left\{-\frac{1}{2 \sigma^2}
(\vec f^T \vec f) [ c - c_0 ]^2\right\}
\qquad.
\end{equation}
One may check that for $\vec f^T \vec f \gg \sigma^2$
it is allowed to shift the integral boundaries to
$+/-$ infinity with affecting the value of the integrand up to a small
error only.
As a matter of fact for the chosen model parameters of $N$=3, $x_i$=$i$,
$T$=1, c=1 and $\sigma$=0.1 the
error is in the order of $10^{-7}$ of the correct integral.
Notice that this is almost the same for every $T$ in between 0 and
infinity.
We finally get
\begin{equation}
p ( T | D , I ) \propto
\frac{ \sqrt{ \vec f^T \vec f } }{ \vec d^T \vec f}
\exp\left\{-\frac{1}{2 \sigma^2}
\left[ \vec d^T \vec d
- \frac{( \vec d^T \vec f )^2}
{\vec f^T \vec f} \right]
\right\}
\qquad.
\end{equation}
A look at the behavior for large and small T gives
\begin{eqnarray}
\lim_{T \rightarrow 0} p ( T | D , I )
& \propto &
\frac{ \sqrt{ N } }{ \sum_i d_i}
\exp\left\{-\frac{1}{2 \sigma^2}
\left[
\vec d^T \vec d
 - \frac{( \sum_i d_i )^2}{ N } \right]
\right\}
\nonumber\\
& \propto & const
\qquad,
\label{womett0}
\end{eqnarray}
\begin{figure}
\includegraphics[height=7cm]{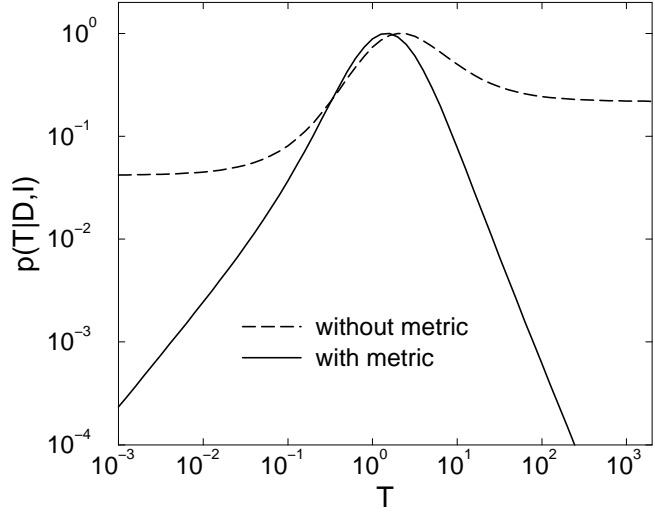}
\caption{
Posterior $p( T | D , I ) $ with (solid line) and without (dashed line)
the Riemannian metric.
Neglection produces non-vanishing tails.
}
\label{figexample}
\end{figure}
\begin{eqnarray}
\lim_{T \rightarrow \infty} p ( T | D , I )
& \propto &
\frac{ \sqrt{ \vec x^T \vec x } }{ \vec d^T \vec x}
\exp\left\{-\frac{1}{2 \sigma^2}
\left[ 
\vec d^T \vec d
 - \frac{( \vec d^T \vec x )^2}
{\vec x^T \vec x} \right]
\right\}
\nonumber\\
& \propto & const
\qquad.
\label{womettinfity}
\end{eqnarray}
Though one has no problem with the lower limit since the integrand is
regular, the non-vanishing posterior distribution for $T \rightarrow \infty$
leads to an expectation value which depends on where the integration limits
are set (see Fig.\ \ref{figexample}).

Now we implement in the calculation the Riemannian metric.
From Eq.\ (\ref{metric}) we get an additional factor $c$, so the
integration over the $c$-dependent parts changes to
\begin{equation}
\frac{1}{n c_0}
\int_{0}^{n c_0}
{\rm d}c \
c \
\exp\left\{-\frac{1}{2 \sigma^2}
(\vec f^T \vec f) [ c - c_0 ]^2\right\}
\qquad.
\end{equation}
Again it is allowed to extend the integration limits to $+/-$ infinity
with only minor error.
The full posterior then gives
\begin{equation}
p ( T | D , I ) \propto
\exp\left\{-\frac{1}{2 \sigma^2}
\left[ \vec d^T \vec d
 - \frac{( \vec d^T \vec f )^2}
{\vec f^T \vec f} \right]
\right\}
\sqrt{
\left(
      \frac{\partial \vec f}{\partial T}^T
      \frac{\partial \vec f}{\partial T}
\right)
-
\left(
\frac{\partial \vec f}{\partial T}^T
\vec f
\right)^2
          / (\vec f^T \vec f)
}
\ .
\label{postwmet}
\end{equation}
What is now the behavior of $p(T|D,I)$ for $T$ approaching 0 and
infinity?
The exponent in Eq.\ (\ref{postwmet}) was already examined in Eqn.\
(\ref{womett0}) and (\ref{womettinfity}) to become constant, so we only
have to look at the square root.
\begin{equation}
\lim_{T \rightarrow 0} 
\sqrt{
\sum_i \left( \frac{x_i}{T+x_i} \right)^4
-
\left[ \sum_i \left( \frac{x_i}{T+x_i} \right)^3 \right]^2
/
\sum_i \left( \frac{x_i}{T+x_i} \right)^2
}
=
\sqrt{N - \frac{N^2}{N}}
= 0
\ ,
\end{equation}
\begin{equation}
\lim_{T \rightarrow \infty} 
\frac{1}{T^2}
\sqrt{
\sum_i \left( \frac{x_i}{1+x_i/T} \right)^4
-
\left[ \sum_i \left( \frac{x_i}{1+x_i/T} \right)^3 \right]^2
/
\sum_i \left( \frac{x_i}{1+x_i/T} \right)^2
}
= 0
\qquad.
\end{equation}
So indeed the square root term which stems from the metric does take care
of zero tails in the posterior!
The nice decrease towards 0 is shown in Fig.\ \ref{figexample} by the
solid line.

\section{Real world problem}

In the problem of determining the electron temperature in an
hydrogen plasma heated
by electron cyclotron resonance, the model function $T$ depends in a quite
complicated way on the temperature $T$:
\begin{equation}
\vec f ( T ) = - \bm{V} ( \bm{R} - \bm{V} )^{-1} \vec x
\qquad.
\end{equation}
Both \bm{V} and \bm{R} are matrices, but only the diagonal matrix
\bm{V} depends on T with entries on the diagonal:
\begin{equation}
V_{ii}
=
\frac{1}{\frac{a_i}{\sqrt{T}}+\frac{1}{b_iT}}
\qquad,
\end{equation}
where $a_i$ and $b_i$ are constants with respect to ion species $i$.
Since the sensitivity of the measurement apparatus is unknown one has to
introduce a linear parameter $c$ in order to relate the data to the
model, i.e.\ Eq.\ (\ref{data}).
Contrary to our simple problem we are not so fortunate to know the
variance $\sigma$ exactly.
The experimentalist can only provide an estimate $\vec s$ of the true errors
$\vec \sigma$ with respect to each other but not on the total scale, so that
we have to introduce an overall multiplication factor $\omega$, with
$\sigma_i = \omega s_i$.
In order to assign a prior to $\omega$ the outlier tolerant approach
\cite{dol99} was chosen:
\begin{equation}
p ( \omega | \alpha , \gamma I )
=
2
\frac{\alpha^\gamma}{\Gamma ( \gamma ) }
\left( \frac{1}{\omega} \right)^{2 \gamma}
\exp \left\{ - \frac{\alpha}{\omega^2} \right\}
\frac{1}{\omega}
\qquad.
\end{equation}
The expectation value of $\omega$ should be one, since the
experimentalist does his estimation according to his best knowledge.
Furthermore, from the characteristics of the measurement process one can
tell that the best guess of $\vec s$ should not deviate by more than 
50\% from the true $\vec \sigma$.
This results in $\alpha=1.28$ and $\gamma=2.0076$.

Now we follow the route explained above to evaluate the expectation
value of $T$.
Again we start by marginalizing $c$ (with the flat prior of Eq.\ (\ref{cprior}))
and $\omega$ without making use of the Riemannian metric.
This gives the posterior in $T$
\begin{equation}
p ( T | D , I ) \propto
\frac{ \sqrt{ \vec {\hat f}^T \vec {\hat f}} }{ \vec {\hat d}^T \vec
{\hat f}}
\left[
\alpha + \frac12 
\left(
\vec{\hat d}^T \vec{\hat d}
-
\frac{(\vec{\hat d}^T \vec{\hat f})^2}{ \vec{\hat f}^T \vec{\hat f}}
\right)
\right]^{- \frac{N}{2}-\gamma-1}
\qquad.
\end{equation}
\begin{figure}[th]
\includegraphics[height=7cm]{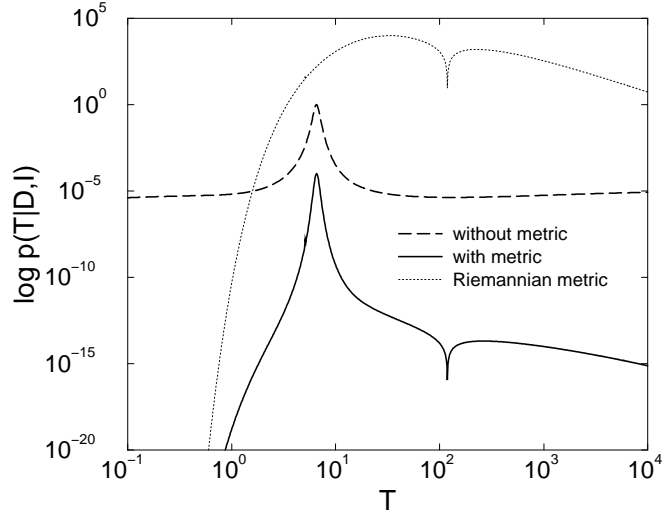}
\caption{
Posterior $p( T | D , I ) $ with (solid line) and without (dashed line)
the Riemannian metric (dotted line).
The incision at T = 118.43 K is a single point which is due to the
parameterization of the physical model.
It does not affect the integrability.
}
\label{figreal}
\end{figure}
For simplicity of notation the hat shall denote that the values have
been divided by the estimated error $\vec s$: $\hat d_i = d_i/s_i$.
The posterior is displayed in Fig.\ \ref{figreal}.
Here we have to face the problem we observed above in the simple example.
Though a non-vanishing tail for $T \rightarrow 0$ is not so harmful, the
increase with $T \rightarrow \infty$ results in a divergence.

Help comes by obeying the correct Riemannian metric.
Then the posterior reads
\begin{equation}
p ( T | D , I ) \propto
\mu(T)
\frac{1}{\sqrt{ \vec {\hat f}^T \vec {\hat f}}}
\left[
\alpha + \frac12 
\left(
\vec{\hat d}^T \vec{\hat d}
-
\frac{(\vec{\hat d}^T \vec{\hat f})^2}{ \vec{\hat f}^T \vec{\hat f}}
\right)
\right]^{- \frac{N}{2}-\gamma-1}
\end{equation}
where $\mu(T)$ is just the metric of Eq.\ (\ref{metric}) without the
terms in $c$ and $\omega$ (marginalized over).
The situation changes completely (see Fig.\ \ref{figreal}) and the
integral becomes feasible now.

\section{Conclusion}
The correct mathematical way to deal with marginalization integrals is to
use the Riemannian metric.
This invariant measure takes care of defining correct infinitesimal
elements to be integrated over.
Since parameterizations of a model may be subjective and vary with the
investigator of a problem, this is the only consistent way to get comparable
answers in probability space.

\section{Acknowledgment}

We like to acknowledge discussions with C.\ Rodriguez.


\begin{thebibliography}{1}

\bibitem{rod89}
C.~Rodriguez, ``The metrics induce by the kullback number,'' in {\em Maximum
  Entropy and Bayesian Methods},  J.~Skilling, ed., Kluwer Academic, Dordrecht,
  1989.

\bibitem{ama85}
S.~Amari, {\em Differential-Geometrical Methods in Statistics},
  Springer-Verlag, Berlin, Heidelberg, 1985.

\bibitem{rod91}
C.~Rodriguez, ``From euclid to entropy,'' in {\em Maximum Entropy and Bayesian
  Methods},  J.~W.~T.~Grandy, ed., Kluwer Academic, Dordrecht, 1991.

\bibitem{dol99}
V.~Dose and W.~von~der Linden, ``Outlier tolerant parameter estimation,'' in
  {\em Maximum Entropy and Bayesian Methods},  \mbox{V. Dose}~et al., ed.,
  Kluwer Academic, Dordrecht, 1999.

\end{thebibliography}
\end{document}